\documentclass[twocolumn]{aastex63}
\usepackage{color, mhchem, here}
\usepackage{bm}
\received{}
\revised{}
\accepted{}
\submitjournal{ApJ}

\shorttitle{Disk photoevaporation and UV pumping}
\shortauthors{Komaki et al.}

\newcommand{\rg}{r_{\rm g}}
\newcommand{\kms}{{\rm km\,s^{-1}}}
\newcommand{\au}{{\rm \,au}}
\newcommand{\braket}[1]{\left(#1\right)}
\newcommand{\gram}{{\rm \, g}}
\newcommand{\Msun}{{\rm \, M_\odot}}
\newcommand{\cm}{{\rm \, cm}}
\newcommand{\erg}{{\rm \, erg}}
\newcommand{\eV}{{\rm \, eV}}
\newcommand{\keV}{{\rm \, keV}}

\newcommand{\Kelvin}{{\rm \,K}}
\newcommand{\Myr}{{\,\rm Myr}}
\newcommand{\yr}{{\,\rm yr}}
\newcommand{\second}{{\rm \, s}}
\newcommand{\figref}[1]{Figure~\ref{#1}}
\newcommand{\tabref}[1]{Table~\ref{#1}}
\newcommand{\eqnref}[1]{Eq.~\ref{#1}}
\newcommand{\secref}[1]{Section~\ref{#1}}

\begin{document}

\title{The effect of ultra-violet photon pumping of H$_2$ in dust-deficient protoplanetary disks}

\author[0000-0002-9995-5223]{Ayano Komaki}
\affiliation{Department of Physics, The University of Tokyo, 7-3-1 Hongo, Bunkyo, Tokyo 113-0033, Japan}
\email{ayano.komaki@phys.s.u-tokyo.ac.jp}

\author[0000-0003-2309-8963]{Rolf Kuiper}
\affiliation{Faculty of Physics, University of Duisburg-Essen, Lotharstra\ss e 1, D-47057 Duisburg, Germany}

\author[0000-0001-7925-238X]{Naoki Yoshida}
\affiliation{Department of Physics, The University of
Tokyo, 7-3-1 Hongo, Bunkyo, Tokyo 113-0033, Japan}
\affiliation{Kavli Institute for the Physics and Mathematics of the Universe (WPI), UT Institute for Advanced Study, The University
of Tokyo, Kashiwa, Chiba 277-8583, Japan}
\affiliation{Research Center for the Early Universe (RESCEU), School of
Science, The University of Tokyo, 7-3-1 Hongo, Bunkyo, Tokyo 113-0033, Japan}



\begin{abstract}
We perform radiation hydrodynamics simulations to study the structure 
and evolution of a photoevaporating protoplanetary disk. 
Ultraviolet and X-ray radiation from the host star heats the disk surface, where \ce{H2} pumping also operates efficiently.
We run a set of simulations with varying the amount of dust grains, or the dust-to-gas mass ratio, which determines the relative importance between photoelectric heating and \ce{H2} pumping.
We show that \ce{H2} pumping and X-ray heating contribute stronger to the mass-loss of the disk if the dust-to-gas mass ratio is $\mathcal{D}\leq10^{-3}$.
The disk mass loss rate decreases with a lower dust amount, but remains around 
$10^{-10-11} M_{\odot} {\rm yr}^{-1}$.
In such dust-deficient disks, \ce{H2} pumping enhances photoevaporation from the inner disk region and shapes the disk mass-loss profile. 
We thus argue that the late-stage disk evolution is affected by the ultra-violet \ce{H2} pumping effect.
The mass-loss rates derived from our simulations can be used in the study of long-term disk evolution.
\end{abstract}



\section{Introduction} \label{sec:intro}
Circumstellar disks around newly born stars evolve to be protoplanetary disks (PPDs) where planets are formed.
Near infrared observations show that the disk fraction in a star-forming region decreases as the age of the central star, suggesting that PPDs disappear in 3--6 Myr \citep[e.g.,][]{Haisch:2001, Meyer:2007, Hernandez:2007, Mamajek:2009, Bayo:2012, Ribas:2014}.

Photoevaporation is suggested to be an important disk dispersal mechanism.
High-energy photons such as far-ultraviolet (FUV; $6\eV \lesssim h\nu < 13.6\eV $), extreme-ultraviolet (EUV; $13.6\eV \lesssim h\nu \lesssim 100\eV$), and X-ray ($100\eV \lesssim h\nu \lesssim 10\keV$) emitted from the central star
heat a large portion of a PPD through various
radiative and photo-chemical processes.
It is generally found that EUV photons are mainly absorbed by abundant neutral hydrogen close to the central star, while FUV and X-ray photons penetrate deep into the disk to drive dense photoevaporative flows \citep[e.g.,][]{RichlingYorke:2000, Ercolano:2009, GortiHollenbach:2009}.
FUV photons are the cause of the photoelectric effect on dust grains in the disk, whereas EUV and X-ray contribute to the heating through ionization of hydrogen, helium, and other heavy elements. 
These processes altogether raise the gas temperature of the disk surface and generate photoevaporative flows.

Previous studies showed that the strength of photoevaporation is determined by a number of physical properties such as metallicity, stellar mass, luminosity and chemical composition  \citep{Ercolano:2009,Nakatani:2018a,Nakatani:2018b,Wolfer:2019,Komaki:2021}.
For example, \cite{Nakatani:2018a} performed radiation hydrodynamics simulations with varying the gas metallicity, 
and showed
that photoevaporation is suppressed in low-metallicity disks because of the low abundance of dust grains. The mass-loss rate and the overall disk evolution
are highly sensitive to the effective
disk heating rate, and thus it is important to perform direct
numerical
simulations in order to 
study the structure of a photoevaporating disk and to derive realistic physical and chemical profiles under a variety of environments.


The disk dust properties are characterised by the dust-to-gas mass ratio and the size distribution.
Since these quantities may be inhomogeneous in time and in space within the disk, the relative importance of the heating processes can also significantly change according to the disk evolution.
Dust grains undergo physical processing during the disk evolution.
Numerical calculations show that small dust grains entrain photoevaporative flows \citep{Hutchison:2016}.
Dust settling may occur in PPDs, as inferred from infrared observations and SED modelling \citep{DAlessio:1999,DAlessio:2006,Grant:2018}.
Dust sedimentation can also significantly reduce the local dust-to-gas mass ratio of the disk surface.
Millimeter observations suggest that the dust mass decreases as the system age increases \citep{Mathews:2012, Ansdell:2016, Pascucci:2016, Ansdell:2017}.
The total dust-to-gas mass ratio of a disk likely varies through its evolutionary phases.

\citet{Gorti:2015} perform a set of one-dimensional disk evolution simulations by incorporating viscous accretion, photoevaporation,
and dust evolution. 
They show that the dust evolution results in the increase of small dust grains, which affects the dust attenuation, photoelectric heating, dust-gas collision, and formation of molecules. 
Clearly, dust grain growth is an important process for disk
evolution, and the amount of dust can critically affect the disk
structure and dispersal process.
Unfortunately, previous studies do not address explicitly how
the reduced amount of dust grains, i.e., low dust-to-gas ratios, 
affect photoevaporation and disk dispersal efficiency.

In this paper, we study the detailed process of photoevaporation and the dispersal of dust-deficient disks.
Following the basic methodology of \cite{Komaki:2021}, we
perform a set of radiation hydrodynamics simulations.
Our simulations incorporate the so-called UV pumping of \ce{H2} molecules (\ce{H2} pumping hereafter), which can be an important heating process in the region where a substantial amount of \ce{H2} molecules are continuously formed and destroyed. 
\ce{H2} pumping is a two-step process; Lyman-Werner (LW) photons in the energy range $11.2\eV \lesssim h\nu < 13.6\eV $ excite \ce{H2} molecules to the electronically excited state, followed by de-excitation to vibrationally excited states of the ground electronic state \citep{TielensHollenbach:1985}. 
Collisional de-excitation to the vibrational ground states 
results in heating of the gas by distributing the de-excitation energy
to the surrounding atoms and molecules.
Note that, while the FUV photoelectric heating rate roughly scales with the local amount of small dust grains, the \ce{H2} pumping operates even in regions with little dust content.
In order to examine how the disk's thermo-chemical structure changes with dust amount, we perform radiation hydrodynamics simulations, for the first time, with varing the dust-to-gas mass ratio of the disk, $\mathcal{D}$, in the range of $10^{-6}$--$10^{-1}$.

\ce{H2} pumping is incorporated to closely examine the effect of \ce{H2} pumping on the disk temperature.
We explain our methods in Section 2.
We show our results in Section 3 and discussions in Section 4.
In Section 5 we summarize.


\section{Numerical simulations}
We perform disk photoevaporation simulations including hydrodynamics, radiative transfer and non-equilibrium chemistry in a coupled and self-consistent manner.
We use the open source code PLUTO \citep{Mignone:2007}
suitably modified for simulations of PPDs.
The details of the implemented physics are found in \citet{Nakatani:2018a,Nakatani:2018b} and \citet{Komaki:2021}.
We assume that the disk is axisymmetric around the rotational axis and adopt the 2D spherical coordinates ($r, \theta$).
We consider three components of the gas velocity, $\bm{v}=(v_r, v_{\theta}, v_{\phi})$.

We first run our fiducial case, where the stellar mass $M_*=1\Msun$ and $\mathcal{D} = 10^{-2}$, with a detailed treatment of \ce{H2} pumping. 
In the case of $\mathcal{D}=10^{-2}$, we calculate the dust temperature in the same way as in \cite{Komaki:2021}, where we calculate and tabulate the dust temperature as a function of the distance from the central star and the column density, based on the results of more costly radiative transfer calculations.
In our runs with other dust-to-gas mass ratios, we solve the dust temperature in a self-consistent manner
by incorporating both the direct and diffuse radiation.
We perform ray-tracing for the direct component and adopt the flux-limited-diffusion (FLD) approximation for the diffused component \citep{Kuiper:2010, KuiperKlessen:2013, Kuiper:2020}.

The governing equations are given by
\begin{gather}
\frac{\partial \rho}{\partial t} + \bm{\nabla} \cdot (\rho \bm{v}) = 0,\\
\frac{\partial (\rho v_{r})}{\partial t} + \nabla\cdot (\rho v_{r} \bm{v}) = - \frac{\partial P}{\partial r} - \rho\frac{GM_{*}}{r^2} + \rho\frac{v_{\theta}^2+v_{\phi}^2}{r},\\
\frac{\partial (\rho v_{\theta})}{\partial t} + \nabla\cdot (\rho v_{\theta} \bm{v}) = - \frac{1}{r}\frac{\partial P}{\partial \theta} - \rho\frac{v_{r}v_{\theta}}{r} + \rho\frac{v_{\phi}^2}{r}\cot\theta,\\
\frac{\partial (\rho v_{\phi})}{\partial t} + \nabla^{l}\cdot(\rho v_{\phi}\bm{v}) = 0,\\
\frac{\partial E}{\partial t} + \nabla \cdot H \bm{v}
= - \rho v_{r}\frac{GM_{*}}{r^2} + \rho(\Gamma - \Lambda),\\
\frac{\partial n_{\ce{H}}y_{i}}{\partial t} + \nabla\cdot (n_{\ce{H}}y_{i}\bm{v}) = n_{\ce{H}}R_{i}.
\end{gather}
The first five equations express fluid equations, and the last one 
coupled with the rate equations expresses the abundance evolution of chemical species.
We use $\rho$, $\bm{v}$, $P$, and $M_{*}$ to represent gas density, velocity, pressure, and stellar mass, respectively.
The fifth equation(Eq.~(5)) is the energy equation where $E$ and $H$ express the total energy and enthalpy, including the kinetic energy per unit volume.
In the equation, $\Gamma$ and $\Lambda$ are the specific heating rate and the specific cooling rate, which means the heating and cooling rate per unit mass.
The azimuthal component of the Euler equation is described in the angular momentum conservation form(Eq.~(6)).
We define $y_{i}$ as the abundance of chemical species $i$,
$n_{\rm H}$ as the number density of elemental hydrogen, 
and $R_{i}$ as the relevant chemical reaction rate.
We follow the chemical reactions tabulated in \citet{Nakatani:2018a, Nakatani:2018b}.
We use the parallel simulation code PLUTO \citep{Mignone:2007} to solve hydrodynamics.


We incorporate EUV/X-ray photoionization heating, FUV photoelectric heating \citep{BakesTielens:1994}, and heating by \ce{H2} photodissociation and by \ce{H2} pumping.
We also incorporate dust-gas collisional cooling \citep{YorkeWelz:1996}, fine-structure cooling of \ion{C}{2} and \ion{O}{1} \citep{HollenbachMcKee:1989, Osterbrock:1989, Santoro:2006}, molecular line cooling of $\ce{H2}$ and $\ce{CO}$ \citep{Galli:1998, Omukai:2010}, hydrogen Lyman $\alpha$ line cooling \citep{Anninos:1997}, and radiative recombination cooling \citep{Spitzer:1978} as cooling sources.
We calculate the \ion{C}{2} and \ion{O}{1} cooling rates assuming the LTE level populations.

The chemical reaction network follows the abundances of eleven chemical species: \ion{H}{1}, \ion{H}{2}, \ce{H-}, \ion{He}{1}, \ce{H2}, \ce{H2+}, \ce{H2^*}, \ce{CO}, \ion{O}{1}, \ion{C}{2} and electrons in the run with $M_*=1\Msun$, $\mathcal{D}=10^{-2}$.
We denote the vibrational excited state of \ce{H2} ($v = 6$) as \ce{H2^*}
and treat as a distinct chemical species \citep{TielensHollenbach:1985}.
\ce{H2} has hundreds of excited states, but the pseudo-level of $v=6$ approximately represents the excited levels altogether, which then reduces the computational cost significantly.
We also use the pseudo-level of \ce{H2^*} to study the effect of \ce{H2} pumping on the disk thermal structure.
We assume that \ion{C}{1} is quickly ionized to become \ion{C}{2} after the dissociation of \ce{CO} \citep{NelsonLanger:1997, RichlingYorke:2000}.


We incorporate \ce{H2^*} to calculate both the excitation rate and the de-excitation rate.
We calculate the \ce{H2} pumping rate, $R_{\rm{pump}}$ as
\[
R_{\rm{pump}}=3.4\times10^{-10}\beta G_0 e^{-2.5A_{\rm{v}}}\second^{-1},
\]
following \cite{TielensHollenbach:1985}.
We define $\beta$, $G_0$ and $A_{\rm{v}}$ as the shielding factor, the intensity of the incident FUV field and the visual extinction respectively.
We determine the shielding factor by \ce{H2} following \cite{DraineBertordi:1996}.
The \ce{H2} molecules in the vibrational excited state de-excite and fluoresce.
We incorporate the collisional de-excitation by \ion{H}{1} or \ce{H2} as the de-excitation reactions.
We also consider the direct dissociation of \ce{H2^*} by FUV radiation and by spontaneous radiation \citep{TielensHollenbach:1985}.
We define $k_{\rm{de}}(\rm{H})$, $k_{\rm{de}}(\rm{H_2})$, $R_{\rm{de}}$, $A(\rm{H_2^*})$ as the reaction rates by the collision with \ion{H}{1}, the collision with \ce{H2}, the direct dissociation by FUV radiation and the spontaneous radiation respectively.
The reaction rates are written as
\begin{equation}
\begin{split}
   k_{\rm{de}}(\rm{H})& \simeq 1.8\times10^{-13}\,\,\cm^3\second^{-1}\\
   &\times\left(\frac{T_{\rm{gas}}}{\Kelvin}\right)^{1/2}\exp\left(-\frac{1000\Kelvin}{T_{\rm{gas}}}\right) 
\end{split}
\end{equation}
\begin{equation}
\begin{split}
    k_{\rm{de}}(\rm{H_2})& \simeq 2.3\times10^{-13}\,\,\cm^3\second^{-1}\\
    &\times\left(\frac{T_{e\rm{gas}}}{\Kelvin}\right)^{1/2}\exp\left(-\frac{1800\Kelvin}{(T_{\rm{gas}}+1200\Kelvin)}\right)
\end{split}
\end{equation}
\begin{gather}
R_{\rm{de}} = 10^{-11}\, \beta \, G_0 e^{-2.5A_{\rm{v}}}\,\,\second^{-1}\\
A(\rm{H_2^*}) = 2.0\times10^{-7}\,\,\second^{-1},
\end{gather}
where $T_{\rm{gas}}$ represents the gas temperature \citep{WangGoodman:2017}.
Each collisional de-excitation process deposits $2.6\eV$ and each de-excitation by direct FUV radiation deposits $0.4\eV$ \citep{TielensHollenbach:1985}.

We explicitly treat \ce{H2^*} as a distinct chemical species. We have run an additional simulation with a simpler
implementation for comparison and testing purpose as follows.
In the fiducial run with $M_{*}=1\Msun$
and $\mathcal{D}=10^{-2}$, we calculate the heating rate by \ce{H2} pumping without explicitly including the excited molecular hydrogen, as in \citet{Rollig:2006}, \citet{Gressel:2020} and \citet{Nakatani:2021}.
Specifically, we calculate the heating rate as follows.
\begin{equation}
\frac{{\rm d}e}{{\rm d}t}= n_{\ce{H2}}\frac{\chi P_{\rm{tot}}\Delta E_{\rm{eff}}}{1+[A_{\rm{eff}}+\chi D_{\rm{eff}}]/[\gamma_{\rm{eff}}n]},
\label{eq:heat_without_H2star}
\end{equation}
where $\chi$ is the FUV radiation intensity in units of the local interstellar radiation field, $P_{\rm{tot}}$ and $n$ are the formation rate of \ce{H2^*} and the particle number density, respectively.
The coefficients $\Delta E_{\rm{eff}},A_{\rm{eff}}, D_{\rm{eff}},\gamma_{\rm{eff}}$ express the average excitation energy, the effective spontaneous emission rate, the effective dissociation rate, and the effective collisional de-excitation rate assuming that molecular hydrogen has two levels, the ground state and the pseudo-excited level. 
We set $P_{\rm{tot}}\Delta E_{\rm{eff}}=9.4\times10^{-22}\erg\second^{-1}$, $A_{\rm{eff}}=1.9\times10^{-6}\second^{-1}$, $D_{\rm{eff}}=4.7\times10^{-10}\second^{-1}$ and $\gamma_{\rm{eff}}=5.4\times10^{-13}\sqrt{T_{\rm{gas}}}\second^{-1}\cm^{-3}$ following \citet{Rollig:2006} and \citet{Gressel:2020}. 
\citet{Rollig:2006} derived Equation (\ref{eq:heat_without_H2star}) by fitting the heating rate calculated in the thermodynamical simulations that consider 100 energy levels of \ce{H2}.
We have tested and checked that using the fitting formula does not change the resulting mass-loss profiles appreciably, while reducing the computational time by $\sim1.5$ times. 
We compare the result further in detail in Section 3.1.
In the following, we shall refer to the heating rate for \ce{H2} pumping (Eq.  [\ref{eq:heat_without_H2star}]) as R{\"o}llig method.

 
\begin{figure*}[htbp]
       \centering
         \includegraphics[width=\linewidth,clip]{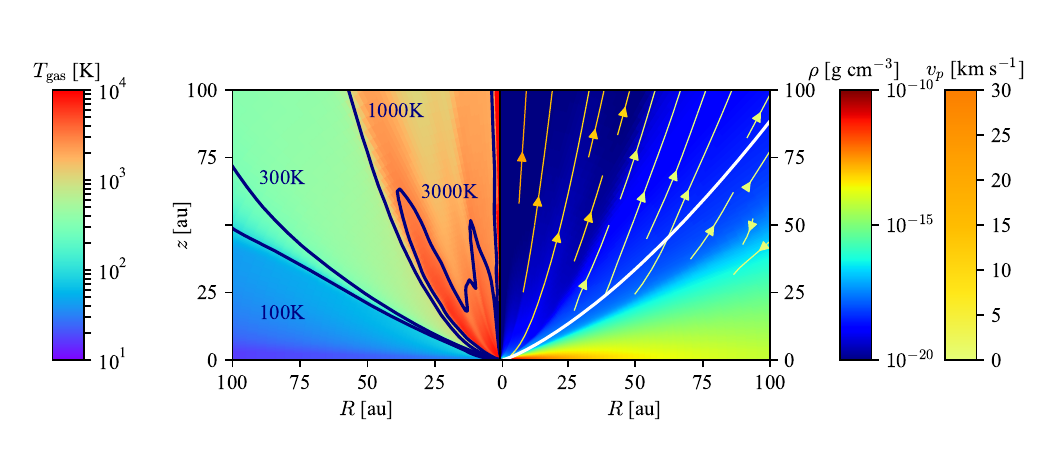}
         \includegraphics[width=\linewidth,clip]{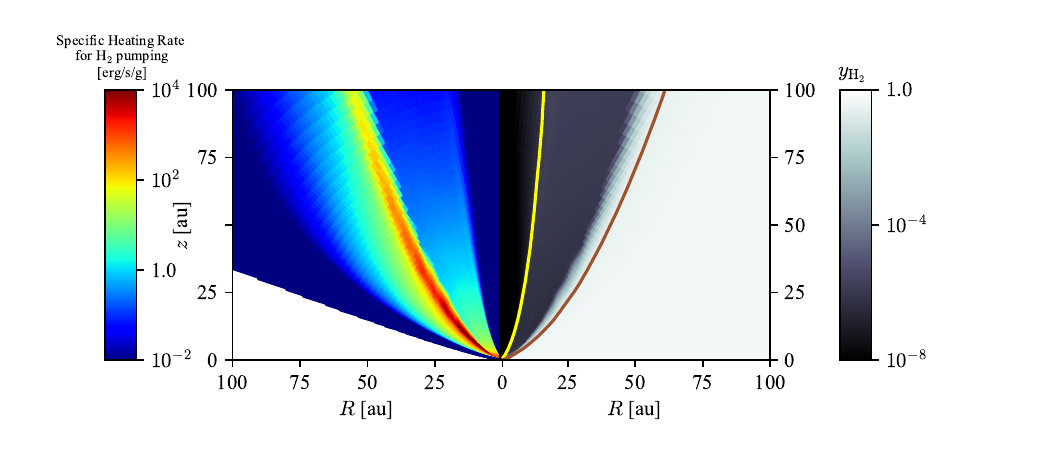}
         \caption{(top) The time-averaged disk structure of the run incorporating \ce{H2} pumping. 
         The color map on the left portion shows the gas temperature, $T_{\rm{gas}}$, and the right portion shows the density distribution, $\rho$. The white line shows the location of the base, which satisfies $N_{\rm{H_2}}=10^{20}\cm^{-2}$. The navy contour lines on the left side represent where the gas temperature is $100\Kelvin$, $300\Kelvin$, $1000\Kelvin$, and $3000\Kelvin$. The streamline represents the poloidal velocity of the gas, which is defined as $v_p=\sqrt{v_r^2+v_{\theta}^2}$.
         For clarity, we omit to plot the streamlines where the velocity is less than $0.1\kms$, which are mostly seen in the optically thick disk region. 
         (bottom) The time-averaged disk chemical structure of the run incorporating \ce{H2} pumping.The color map on the left portion shows the specific heating rate for \ce{H2} pumping. The right portion shows the abundance of \ce{H2}.The right portion also shows the distribution of H-bearing species. The yellow line on the right side represents the ionization front where the abundance of \ion{H}{2} is 0.5, and the brown line indicates the dissociation front where the abundance of \ce{H2} is 0.25.
         }
      \label{fig:snapshots}
\end{figure*}

The size distribution of dust grains is assumed to follow ${\rm d} n(a)\propto a^{-3.5} {\rm d}a$ with $a$ denoting dust radii \citep{DraineLee:1984}.
\cite{BakesTielens:1994} show that small grains are crucial for efficient photoelectric heating.
In practice, dust dynamics such as sedimentation, radial drift and/or turbulent mixing can change the dust distribution across the disk.
However, for simplicity, we set the dust-to-gas mass ratio to constant throughout the computational domain. 
The underlying assumption is that the dust grains responsible for photoelectric heating are well mixed within the gas so that we treat the mixture of dust and gas as a single fluid. 
We perform a set of simulations with different constant values of the dust-to-gas mass ratio.
Note that the dust amount near the disk surface, where photoevaporative flows are launched, critically determines the disk mass-loss rate.
We perform disk evolution simulations with varying the dust-to-gas mass ratio of the disk $\mathcal{D}$
in the range of $10^{-6}$--$10^{-1}$ as a parameter to examine how the effective heating and cooling processes vary with $\mathcal{D}$.
When we adopt lower $\mathcal{D}$, we assume that the dust grains of all the sizes are depleted equally. 
Since photoelectric heating is largely caused by dust grains smaller than $\sim 20\mu$m, the value of $\mathcal{D}$ in the present study is meant to reflect the amount of small dust grains that contributes to the photoelectric heating.
In our simulations, the dust opacity, photoelectric heating rate, dust-gas collisional cooling rate, the rate of \ce{H2} formation catalyzed by grains are scaled with $(\mathcal{D}/0.01)$. 
Our fiducial dust-to-gas mass ratio is $\mathcal{D}=10^{-2}$, which corresponds to the local ISM value.
The gas-phase elemental abundance of carbon is set to $y_{\rm{C}}=0.927\times10^{-4}$ and that of oxygen to $y_{\rm{O}}=3.568\times10^{-4}$ \citep{Pollack:1994}.
We assume that $y_{\rm C}$ and $y_{\rm O}$ are constant even when we set different $\mathcal{D}$.


We take into account FUV, EUV and X-ray radiations as the high-energy radiation from the central star.
As in our previous paper \citet{Komaki:2021}, we choose the FUV, EUV and X-ray luminosities, $\phi_{\rm{EUV}}$, $L_{\rm{FUV}}$, $L_{\rm{X}}$, following Table~1 in \cite{GortiHollenbach:2009}, which shows typical $1\Myr$ pre-main sequence star luminosities.

We define the computational domain of the polar angle to [0, $\pi /2$], and the radial coordinate to [$0.1\rg$, $20\rg$].
The gravitational radius $\rg$ for a fully ionized gas with $T_{\ce{gas}} = 10^4\Kelvin$ 
is
\[
\rg = \frac{GM_*}{(10\kms)^2} \simeq 8.87 \au \braket{\frac{M_*}{1\Msun}}, 
\]
where $G$ is the gravitational constant.
The gravitational radius sets the estimate of the effective boundary where the photoevaporative flows can escape out of the potential well of the host star.
However hydrodynamics simulations show that the photoevaporative flows are also excited inside the gravitational radius from the critical radius $\sim 0.2 \rg$ \citep{Liffman:2003, Font:2004}.
Motivated by findings, we set the computational range to include the region inside the graviational radius.
We use $\rg$ as a reference physical length scale in this paper.

We start our simulations setting the initial disk condition to be a 1 million year system, when the mass loss by accretion becomes lower \citep{Clarke:2001, Alexander:2006, Owen:2010, Suzuki:2016, Kunitomo:2020}.
We assume that the disk mass, $M_{\rm{disk}}$, is 3$\%$ of the central stellar mass \citep{AndrewsWilliams:2005}.
We assume the initial surface density profile, $\Sigma$ and the initial temperature distribution, $T_{\rm{ini}}$, as \cite{Komaki:2021}, to be
\[
\begin{split}
\Sigma (R) &= 27.1\gram\cm^{-2}\times (R/\rg)^{-1}\\
T_{\rm{ini}} &= 100\Kelvin\times (R/0.1\rg)^{-1/2}.
\end{split}
\]

We run the simulations for $8.4\times 10^{3}(M_*/\Msun)$ yr.
This is 10 times of the gas Kepler rotational period around the central star at $r=\rg$ and this is long enough for the time-averaged mass-loss to converge.

\section{Results}   \label{sec:results}
In this section, we present the results of our simulations.
We first describe the effect of \ce{H2} pumping on the temperature and the chemical structure in the run with $M_*=1\Msun$ and $\mathcal{D}=10^{-2}$ in \secref{sec:result1}.
We then calculate the mass-loss rate of this run in \secref{sec:result2}.
Finally in \secref{sec:result4}, we describe how the disk thermal structure changes when varying the dust-to-gas mass ratio. 

\subsection{Thermochemical Structure}\label{sec:result1}

\figref{fig:snapshots} shows the time-averaged snapshot of the simulation  with $M_*=1\Msun$ and $\mathcal{D}=10^{-2}$ and with incorporating \ce{H2^*} and \ce{H2} pumping.
To make this plot, we average the outputs from $840\yr$ to $8400\yr$ to avoid the initial transient and to smooth out (temporarily) fluctuating features.
Photoevaporative flows are launched from the surface of the disk, where the column density satisfies $N_{\rm{H_2}}\approx10^{20}\cm^{-2}$
(white line in the top panel).
We define this surface as the base of the photoevaporative flows.

EUV radiation heats the gas near the central star and at high latitudes,
shaping the polar \ion{H}{2} region where $y_{\rm{HII}}>0.5$.
The outgoing gas cools by adiabatic cooling
and has a temperature of $T_{\rm{gas}}\sim3000\Kelvin$.

The bottom portion of \figref{fig:snapshots} shows the \ce{H2} pumping heating rate $\Gamma_{{\rm H}_2^*}$ and also the molecular fraction $y_{\rm H2}$.
The \ce{H2^*} is most abundant near the dissociation front
with a number fraction $\sim0.04$, where $\Gamma_{{\rm H}_2^*}$
reaches $\sim10^{4}\erg\second^{-1}\gram^{-1}$.
We note that the \ce{H2} dissociation front locates at 
a lower polar angle than the base.
The temperature of the photoevaporative flows is determined by the balance between FUV photoelectric heating and \ce{H2} cooling in the neutral region.
At the base of the outflow, \ce{OI} cooling is the main cooling process.

In the dense inner region at $r\lesssim 30\au$,
\ce{H2^*} molecules get de-excited mainly by collisions with neutral hydrogen atoms.
There, the gas is largely heated by this \ce{H2} pumping process.
The critical density for \ce{H2^*} spontaneous radiation is $\rho_{\rm c} \simeq5.0\times10^{-20}\gram\cm^{-3}$ assuming $T_{\rm{gas}}=3000\Kelvin$.
In our simulation, the density at $r\gtrsim 30\au$ is lower than 
$\rho_{\rm c}$, and thus  
the spontaneous radiation primarily de-excites the \ce{H2^*} molecules there, and the resulting gas heating rate is small.
We note that, in our simulations, the timescale for \ce{H2^*} de-excitation is shorter than the local hydrodynamical timescale by a factor of $\sim 10^{2}$, 
so the \ce{H2^*} molecules almost immediately get de-excited at the location where they are excited by the incident radiation.



We have run an additional simulation with adopting R{\"o}llig method, which does not explicitly treat \ce{H2^*} molecules. 
By closely examining the outputs of the two simulations, 
we find that the disk structure
is essentially the same in regions where the gas temperature is $300\Kelvin\leq T_{\textrm{gas}}\leq 3000\Kelvin$.
Such regions are found typically in the vicinity of the disk surface as shown in \figref{fig:snapshots}, where strong photoevaporative flows are driven. Noticeable differences are found
in low temperature regions with $T_{\textrm{gas}}\leq 300\Kelvin$,
where radiative heating is unimportant. 
We thus conclude that our main results on disk photoevaporation are not significantly affected by the choice of the two calculation methods
of \ce{H2} pumping.



\begin{figure*}
       \centering
         \includegraphics[width=\linewidth,clip]{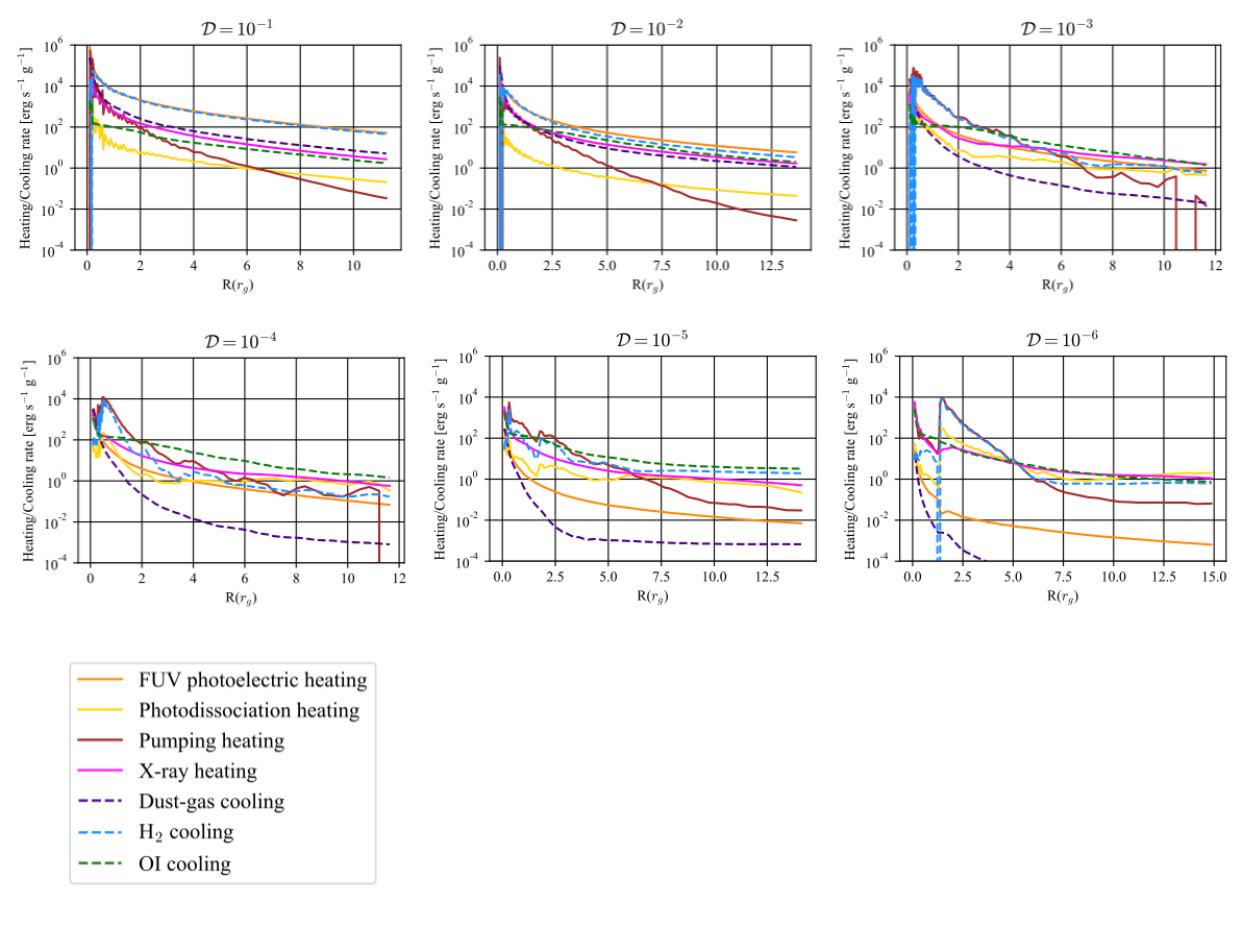}
         \caption{The time-averaged specific heating and cooling rates 
         measured at the base in our runs with different $\mathcal{D}$. The solid lines show FUV photoelectric heating (orange), \ce{H2} photodissociation heating (yellow), \ce{H2} pumping heating (brown), and X-ray heating (magenta). The dotted lines are for dust-gas collisional cooling (purple), \ce{H2} cooling (cyan), and \ion{O}{1} cooling (green).}
         \label{fig:HeatingCoolingDGMR}
\end{figure*}

\subsection{The Mass-loss Rate}\label{sec:result2}
The disk gas is heated by the radiation from the central star, 
and photoevaporateive flows are launched from the disk.
The disk mass-loss rate is determined by the heating process around the base.
In our fiducial run,
FUV photoelectric heating is the main heating source at the base, 
and its rate is greater than the heating rate of \ce{H2} pumping by a factor of $\sim 10^3$ throughout the base.
The resulting temperature distribution is almost the same as in the run without \ce{H2} pumping at the base, indicating that the photoevaporative flows are driven mainly by FUV photoelectric heating.

We use the simulation outputs to calculate the mass-loss rate 
\begin{equation}
\dot{M} = \int_{S, \eta >0} \rho \bm{v}\cdot d\bm{S}, \label{eq:massloss}
\end{equation}
where $d\bm{S}$ represents the spherical surface unit area at $r=100\au$.
We define $\eta$ as the Bernoulli function given by
\[
\eta = \frac{1}{2}v_{p}^{2}+\frac{1}{2}v_{\phi}^{2}+\frac{\gamma}{\gamma -1}c_{\rm{s}}^{2} - \frac{GM_{*}}{r}.
\]
We define $v_p=\sqrt{v_r^2+v_{\theta}^2}$ as the poloidal velocity, $\gamma$ as the specific heat ratio, and $c_{\rm{s}}$ as the sound velocity.
We assume that only the gas satisfying $\eta>0$ has sufficient mechanical energy to escape from the system eventually.

The mass-loss rate slightly fluctuates in time mainly because of spurious reflection of subsonic gas at the outer computational boundary.
The averaged mass-loss rate (from $840\yr$ to $8400\yr$) is
\[
\dot{M}=2.7\times 10^{-9} \Msun\yr^{-1}.
\]
The magnitude of the fluctuation is from $-13\%$ to $+10\%$ compared to the averaged value.
The mass-loss rate with \ce{H2} pumping is almost the same within the fluctuation as that without \ce{H2} pumping ($\dot{M}=2.6\times10^{-9}\Msun\yr^{-1}$).
This is again because \ce{H2} pumping heats the gas mainly in the vicinity of the \ce{H2} dissociation front and does not affect the temperature structure in the photoelectric-heated region.
Hence, the effect of \ce{H2} pumping on driving photoevaporative flows is limited if \ce{H2}-rich flow is excited by photoelectric heating.

We also calculate the mass loss rate for the simulation with  R{\"o}llig method.
The obtained time-averaged $\dot{M}$ differs only by $4\%$ from our fiducial case. This is consistent with the already discussed finding that the accurate treatment of \ce{H2} pumping 
is important only in low-temperature, neutral regions in the disk.
In the following sections, we use simulations that include \ce{H2^*} molecules explicitly.

\subsection{Dependence on the Dust-to-Gas Mass Ratio}\label{sec:result4}
\begin{figure}
       \centering
         \includegraphics[width=\linewidth,clip]{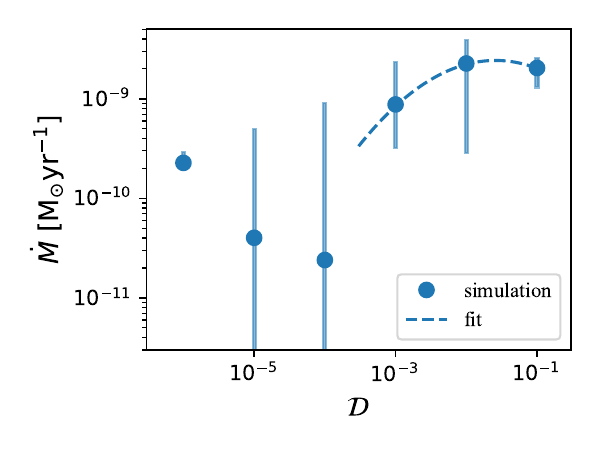}
         \caption{The time-averaged mass-loss rate for each dust-to-gas mass ratio. The blue dots represent the simulation results. The bars indicate the range of typical variation of the instantaneous rate without time-averaging.
          The dotted line is a functional fit we derive for $\mathcal{D}=10^{-3}-10^{-1}$
          where the main disk heating process is photo-electric heating.}
         \label{MassLossDGMR}
\end{figure}

We perform a series of photoevaporation simulations with varying the dust-to-gas mass ratio in the range 
$\mathcal{D}=10^{-6}$--$10^{-1}$.
\figref{fig:HeatingCoolingDGMR} shows the specific heating and cooling rates at the base in each simulation.
In the case of $\mathcal{D}=10^{-1}$, since the dust optical depth is high, FUV photons cannot deeply penetrate into the disk compared to the fiducial case ($\mathcal{D} = 10^{-2}$).
Photoevaporation is caused by FUV photoelectric heating at $N_{\rm{H_2}}=10^{19}\cm^{-2}$, which is one order of magnitude lower than the $\mathcal{D}=10^{-2}$ case. 
(We define this surface as the base for this run.)
The mass-loss rate decreases because of the lower base density (\figref{MassLossDGMR}). 

The FUV photoelectric heating rate approximately follows $\Gamma_{\rm{FUV}}\propto \mathcal{D}$ at the base.
For $\mathcal{D}=10^{-3}$, it gets lower to be comparable to heating rates for X-ray heating and \ce{H2} pumping.
At the outflow base, the main cooling process is \ce{OI} cooling for low dust content with $\mathcal{D}\leq10^{-3}$.
In our simulations, we do not vary the gas metallicity when we decrease the dust-to-gas mass ratio, and thus the relative importance of gas-phase metals increases for lower $\mathcal{D}\leq10^{-3}$.
The temperature of the \ce{H2} region can not get sufficiently high to have $\eta > 0$. It is in contrast to the fiducial run where molecular flow is observed. The mass-loss rate is lower than that of the fiducial run accordingly. 

For $\mathcal{D}=10^{-4}$--$10^{-6}$
, the FUV photoelectric heating is ineffective, and the thermochemical structure differs from the higher-$\mathcal{D}$ runs.
EUV photons heat the gas at high latitudes, whereas
X-ray heating is dominant in the \ion{H}{1} region
instead of FUV photoelectric heating. 
The gas temperature is lowered to $300\Kelvin$ by adiabatic cooling and \ion{O}{1} line emission there.
\ce{H2} pumping is generally dominant only near the dissociation front, and the other \ce{H2} region is heated by X-ray up to $N_{\rm H} \sim 10^{22} \cm^{-2}$. 
\figref{fig:HeatingCoolingDGMR} shows the specific heating and cooling rates along the base.  
The relative contributions of the heating processes vary with the distance from the host star. 
\ce{H2} pumping and X-ray are dominant at the radii inner and outer than $5$--$7\rg$, respectively. The inner regions are favorable for \ce{H2} pumping since the density exceeds the critical value. 
The gas temperature is $200$--$300\Kelvin$ in the molecular region at $r=10\rg$. 
It is lower by a factor of two than would have been achieved by photoelectric heating. 
The same is true at other radii, and thus the $\eta = 0$ boundary is present at two orders of magnitude smaller column density, $N_{\rm{H_2}}=10^{18}\cm^{-2}$
, than the $\mathcal{D}=10^{-2}$ case. 
The flow accordingly has a low density. It results in the significant decrease in the mass-loss rate compared to the $\mathcal{D} \geq 10^{-3}$ runs (see \figref{MassLossDGMR}).

\deleted{
As $\mathcal{D}$ decreases, the photoelectric heating gets weaker.
Smaller $\mathcal{D}$ also implies a lower rate of \ce{H2} formation on grain surfaces.
However, in all the runs, \ce{H2} is mainly provided by advection at the base.
The charge exchange reaction
\begin{equation}
\ce{H2^+}+\ce{H}\ce{->} \ce{H2}+\ce{H+}
\label{H2*destruction}
\end{equation}
dominates the \ce{H2} formation process, and the \ce{H2} abundance is independent of $\mathcal{D}$.
This makes the heating rate 
$\Gamma_{{\rm H}_2^*}$
to remain almost constant, and thus \ce{H2} pumping contributes to the disk heating in all the range of $\mathcal{D}$ of our calculations.
}

\deleted{To distinguish the effects of the EUV ionization and \ce{H2} pumping, we perform a simulation which includes EUV ionization as the only heating process with a $\mathcal{D}=10^{-6}$ case.
The photoevaporative winds are driven by EUV heating and the mass-loss rate is $\dot{M}=5.1\times10^{-11}\Msun\yr^{-1}$, which is one magnitude lower than the fiducial $\mathcal{D}=10^{-6}$ case.
The low mass-loss rate suggest that EUV radiation does not contribute to photoevaporation.}
In order to extract the contribution of EUV-driven flows from the total mass-loss rates in the low-$\mathcal{D}$ cases, 
we count only the mass flux of the ionized outflows ($y_{\rm{HII}}\geq0.5$) using \eqnref{eq:massloss}.
The mass-loss rates of the ionized gas are found to be roughly an order of magnitude smaller than the total, meaning that the contribution of EUV-driven flows is limited. 
This small contribution is mainly due to the fact that \ce{H2} pumping increases the local scale height of the neutral gas at the innermost radii, and the vertically extended gas shields EUV that would otherwise have reached the outer radii. 
As a result, the \ion{H}{2} region is confined within the conical volume at low polar angle. 




\deleted{
Because of the low abundance of dust grains in the low $\mathcal{D}$ cases, the dust-gas collisional heat transfer is inefficient, and the temperature difference between the disk region and the outflow is large.
The gas is mainly heated through dust-gas collisional heat transfer in the disk region, below the base.
The gas temperature and the scale height becomes low consequently.
}


The total mass-loss rates are measured in the same manner as in \secref{sec:result2}.

The mass-loss rate is approximated as
\begin{equation}
    \dot{M}\sim 6.4\times10^{-0.23(\log_{10}\mathcal{D})^2 - 0.73(\log_{10}\mathcal{D})-10}\Msun\yr^{-1}
\label{eq:MassLoss}
\end{equation}
in the cases with $\mathcal{D}\geq10^{-3}$.
In the case with $\mathcal{D}=10^{-4}, 10^{-5}$, the gas temperature is lower than the dust-abundant cases by the factor of $\sim2$, the mass-loss rate becomes low accordingly.
In the case with $\mathcal{D}=10^{-6}$, the gas is not heated by dust-gas collisional heat transfer at the midplane because of the poor dust abundance.
The scale height becomes low and the gas gathers in the low-latitude region consequently.
The base density and the mass-loss rate becomes larger.
Since the abundance of \ce{H2} does not strongly depend on the amount of small dust grains, we expect that the mass-loss rate does not vary from the case with $\mathcal{D}=10^{-6}$ towards even lower values of $\mathcal{D}$.

We note that the actual spectrum includes emission line features in addition to the continuum FUV field.
Observations towards T Tauri stars suggest that hydrogen Lyman-$\alpha$ and \ion{C}{4} line photons cause \ce{H2} pumping \citep[e.g.,][]{Herczeg:2002, Herczeg:2004, Herczeg:2006, Yang:2011}.
These photons have energies lower than $11.2\eV$ but can effectively pump
\ce{H2}$^*$ molecules in the vibrationally excited states. 
Hence strong emission lines can increase the direct photodissociation rate $R_{\rm{de}}$ given by Equation (9).
Since the direct dissociation rate and the associated heating rate are proportional to the FUV flux ($G_0$), we can examine the pumping effect by increasing $R_{\rm de}$ or equivalently $G_0$. To this end, we have performed a test calculation for the case of $\mathcal{D}=10^{-3}$ with
a ten times larger $R_{\rm de}$.
The result shows that the enhanced direct dissociation of \ce{H2^*} does not contribute significantly to the {\it heating rate} because
de-excitation of \ce{H2^*} by collisions with hydrogen atoms is dominant in the inner disk region. The outer part is primarily heated by X-ray photons (Figure 2), and the boosted pumping is not important there.
Lyman-$\alpha$ and \ion{C}{4} photons may affect the abundance of \ce{H2^*} in the vicinity of the central star, but do not affect the gas temperature nor the total mass-loss rate.
In our future work, we will incorporate detailed radiation transfer with realistic spectra with emission lines
for direct comparison with spectral observations.

\cite{Clarke:2001} suggest that the disk dispersal is driven by a combination of accretion and photoevaporation.
In their simulation, the disk gas is dispersed quickly after a gap opens in the disk, and photoevaporation remains effective at the outer region.
\cite{Kunitomo:2020} use a set of disk evolution simulations which incorporate accretion, magnetohydrodynamic (MHD) winds and photoevaporation, to study the respective contribution to disk dispersal.
It is necessary to perform disk evolution simulations
with all the relevant processes implemented accurately, in order to compare the numerical results with observations.
Our simulations allow us to use the detailed structure of evaporating disks for the study of long-term disk evolution,   
and thus enable comparison of the disk dispersal time with observationally inferred disk lifetimes \citep{Komaki:2023}.

\cite{Nakatani:2018b} investigated the dependence of the mass-loss rate on the gas metallicity $Z$.
They varied $Z$ with setting the dust-to-gas mass ratio as $\mathcal{D}=0.01\times(Z/Z_{\odot})$. 
The gas-phase elemental abundances of metals are also scaled in proportion to $(Z/Z_{\odot})$. 
On the other hand, we fix the gas-phase elemental abundances of carbon and oxygen constant regardless of $\mathcal{D}$ in the present study.
Our run with $\mathcal{D}=10^{-2}$ corresponds to the $Z/Z_{\odot}=1$ case of \cite{Nakatani:2018b},
which assumes $\sim10$ times higher luminosities for FUV and EUV, and $\sim 2.5$ times higher X-ray luminosity. The overall stronger radiation yields a higher mass-loss rates in all the metallicity range. The difference between the low-$Z$ runs of \cite{Nakatani:2018b} and our low-$\mathcal{D}$ runs appears in the major cooling process. For example, with small $\mathcal{D}$, \ion{O}{1} line cooling is more effective than dust-gas collisional cooling, which has not been observed in \cite{Nakatani:2018b}.

\cite{WangGoodman:2017} performed hydrodynamics simulations with the same disk mass, the same central stellar mass and the same stellar luminosities as in the present study.
They assume that the disk size is $100\au$ and the abundance of small dust grains is fixed to be $7\times10^{-5}$.
The corresponding total dust-to-gas mass ratio is $\mathcal{D}=2.2\times10^{-3}$ by assuming the MRN dust size distribution.
In order to compare their results directly with our simulations, we estimate the mass-loss rate by using \eqnref{eq:MassLoss}, to obtain
\[
\dot{M}=1.7\times10^{-9}\Msun\yr^{-1},
\]
which is consistent with the result of \cite{WangGoodman:2017}.
In their simulation, the disk surface is heated to $T_{\rm{gas}}>10^4\Kelvin$ by EUV photons.
They assume that each EUV photon has energy of $25\eV$, while we consider a spectral shape of high-energy radiation.
Photons with higher energy tend to penetrate deeper into the gas and give higher energy to the gas.
This leads to the distinct thermal structure with the broader \ion{H}{2} region.

\section{Discussion} 



While the results presented in the previous section have a variety of interesting implications, there are still uncertainties originating from the model assumptions.  
In this section, we further discuss possible variations of our results and show caveats.

\subsection{Dust Evolution}
It is expected that dust sedimentation
causes the local dust-to-gas mass ratio to vary in the vertical direction.
Dust grains can also drift radially 
toward the inner region.
In the present study,
we set the dust-to-gas mass ratio $\mathcal{D}$ 
as a convenient parameter to quantify and examine the effect of dust abundance on disk photoevaporation.
We have shown that the mass-loss rate and the mass-loss profile change sensitively depending on 
$\mathcal{D}$. 

Infrared and millimeter observations suggest that PPDs contain dust grains with various sizes \citep{Acke:2004, Pascual:2016, Davies:2018}.
The dust size distribution and its variation have considerable effects on disk heating and the resulting mass-loss rate.
Small grains have large surface areas per mass, 
and thus contribute predominantly to the net photoelectric heating. 
If a PPD achieves a state deficient in small dust grains,
it is less susceptible to FUV photoelectric heating.

\cite{Hutchison:2016} show that the dust size differs between wind regions and the disk because small dust grains samaller than $0.3\ \rm{\mu m}$ diameter are entrained by photoevaporating flows and the mass-loss rate is two magnitudes smaller than the gas mass-loss rate.
We estimate the effect of dust entrainment in the case where the amount of dust grains decreases for $\sim 3\Myr$.
If we assume a mass loss rate of $\dot{M}\sim10^{-9}\Msun {\rm yr}^{-1}$ (Figure 3), the disk gas mass decreases by $10\%$ by photoevaporation.
If only dust grains smaller than $0.3\ \rm{\mu m}$ are entrained, the sum of dust surface will also decrease by $10\%$.
Then the photoelectric heating rate also decreases by $10\%$ because it is proportional to the dust surface area.
Since the disk lifetime is around a few million years, it is unlikely that the dust transportation by photoevaporative flows affect the disk dispersal significantly more than the above estimate. 

However, mixing of gas and dust near the base may be a complicated process, and thus ideally detailed hydrodynamics simulations with explicit treatment of dust size evolution and dynamical gas-dust (de)coupling would be needed to fully address this issue.

\subsection{X-ray Luminosity Dependence}
%
Recent observations suggest that EUV, FUV, X-ray luminosities differ considerably even among the same spectral-type stars \citep{Gullbring:1998,Flaccomio:2003,France:2012,Vidotto:2014,France:2018}.
\cite{Kunitomo:2021} consider the variation of the emissivity of the high-energy radiation together with
stellar evolution.
For example, for a central star with mass $M_*\geq1.5\Msun$, the X-ray luminosity decreases by a factor of $\sim 10^{4}$ at the age of $1$--$10\Myr$.
To examine the impact of X-ray radiation, we perform an additional set of photoevaporation simulations with varying X-ray luminosities to $L_{\rm{X}}=2.51\times10^{29}\erg\second^{-1}$ and $L_{\rm{X}}=2.51\times10^{31}\erg\second^{-1}$, which are 0.1 and 10 times our fiducial value, $L_{\rm{X,f}}$, respectively.
We keep the stellar mass to $M_*=1\Msun$ and the dust-to-gas mass ratio to $\mathcal{D}=10^{-2}, 10^{-6}$.

\begin{table}[]
    \centering
    \begin{tabular}{ccc}
        $\mathcal{D}$ & $L_{\rm{X}}$ [$\erg\second^{-1}$] & $\dot{M}$ [$10^{-9}\Msun\yr^{-1}$]\\ \hline \hline
        $10^{-2}$ & $2.5\times10^{31}$ & $5.2$\\
        $10^{-2}$ & $2.5\times10^{30}$ & $2.7$\\
        $10^{-2}$ & $2.5\times10^{29}$ & $2.5$\\ \hline
        $10^{-6}$ & $2.5\times10^{31}$ & $1.6$\\
        $10^{-6}$ & $2.5\times10^{30}$ & $0.62$\\
        $10^{-6}$ & $2.5\times10^{29}$ & $0.5$\\ \hline
    \end{tabular}
    \caption{The mass-loss rates in the runs with different $\mathcal{D}$ and $L_{\rm{X}}$.}
    \label{tab:Xraydependence}
\end{table}
The resulting mass-loss rates are listed in \tabref{tab:Xraydependence}.
In the low X-ray cases, the gas is heated by both \ce{H2} pumping and X-ray radiation.
As the X-ray luminosity decreases, the gas temperature in the neutral region becomes lower by the factor of $\sim3$.
This sensitivity indicates that the mass-loss rate is dependent on X-ray luminosity.

\subsection{Other effects}
In our simulations, we set the FUV luminosity without considering the detailed spectral shape of the central star.
Realistic stellar spectrum should include emission lines such as Ly$\alpha$ and \ion{C}{4}.
\cite{Schindhelm:2012} have shown that $\sim81\%$ of the stellar radiation in the FUV range is contributed by hydrogen Ly$\alpha$ photons.
In order to examine the effect of the "boost" of FUV radiation by the line emission, we perform a simulation with 10 times higher $L_{\textrm{FUV}}$.
The result shows that $\dot{M}$ is enhanced by a factor of 3 than the fiducial case, which clearly suggests that the resulting $\dot{M}$ can be higher if 
a realistic spectrum with Ly$\alpha$ radiation is considered.

Finally, we discuss the effect of \ce{H2} pumping
in an extremely metal/dust-poor disk.
An interesting question is how
the \ce{H2} pumping affects the dispersal of disks that consists of 
only hydrogen and helium.
Our simulation with $\mathcal{D}=10^{-6}$ can be regarded as almost a metal-free case in the early universe (\figref{fig:HeatingCoolingDGMR}).
The result shows that the disk around a low-mass star is heated by \ce{H2} pumping and X-ray radiation and loses its gas mass by photoevaporation at a rate of $\sim 2.2 \times 10^{-10} \Msun\yr^{-1}$.
In our future study (Komaki et al., in preparation), we examine the structure, stability and the dispersal of zero-metallicity disks in the context of the formation of the first stars.

\deleted{
\subsection{Comparison with Previous Simulations}
\cite{WangGoodman:2017} performed hydrodynamics simulations with the same disk mass, the same central stellar mass and the same stellar luminosities as ours.
They assume that the disk size is $100\au$ in radius and the dust-to-gas mass ratio is $7\times10^{-5}$.
In order to compare their result directly with our simulations, we calculate the mass-loss rate in the range $r<100\au$ in our simulation with $M_*=1\Msun$, $\mathcal{D}=10^{-4}$.
We obtain
\[
\dot{M}=6.92\times10^{-10}\Msun\yr^{-1},
\]
which is smaller than that of \cite{WangGoodman:2017} by a factor of $\sim3.6$.
In their simulation, the disk surface is heated to $T_{\rm{gas}}>10^4\Kelvin$ by EUV photoionization heating.
The \ion{H}{2} region extends to the wind base, while in our simulation the \ion{H}{2} region is confined to the region at high latitudes.
The evaporating gas moves faster and results in a higher mass-loss rate.


\cite{WangGoodman:2017} derive the time-averaged mass loss rate for the first $500\yr$ after the initiation of radiative-transfer calculation.
We calculate averages over a sufficiently long period of 4500 yrs starting from $t=1500\yr$, in order to avoid including initial transient features and also to consider only the phase when the disk is in a quasi-steady state.
We have found that the mass-loss rate is generally higher at the beginning of our simulation.}

\section{summary}
Recent millimeter observations suggest that the dust amount in a PPD decreases with disk evolution. 
The gas mass variation still remains uncertain, but the dust abundance and its long-term variation 
may be important in the process of disk photoevaporation, and thereby planet formation through disk dispersal.
In the present paper, we have performed a set of
radiation-hydrodynamics simulations with varying the dust-to-gas mass ratio to quantify the effect of \ce{H2} pumping on the photoevaporation of dust-deficient disks.
In the standard case with $M_*=1\Msun$ and $\mathcal{D}=10^{-2}$, \ce{H2} pumping heats the evaporating gas driven by FUV photoelectric heating. \ce{H2} pumping does not directly contribute to the total mass loss but affects the temperature structure of the outflow.
However, we have also shown that \ce{H2} pumping heating is effective at the inner region within several gravitational radii where the density is higher than the critical density for collisional de-excitation of \ce{H2} molecules.


We have found that the major  heating process varies with the dust-to-gas mass ratio of the disk.
In relatively dust-rich cases with $\mathcal{D}\geq 10^{-3}$,
the disks are heated mainly by the FUV photoelectric process, and the outer molecular region evaporates efficiently.
In the other, dust-deficient disks with $\mathcal{D}\leq 10^{-3}$, the dominant heating process is caused by \ce{H2} pumping and X-ray radiation. 
The derived mass-loss rate as a function of $\mathcal{D}$ is given by \eqnref{eq:MassLoss} for the cases of $\mathcal{D}\geq10^{-3}$.
The mass-loss rate is relatively small at $\mathcal{D} < 10^{-3}$ where the FUV photoelectric heating is inefficient.

Since the dust amount can vary on a long timescale as the disk evolves, the temperature structure, the mass-loss profile, and the chemical composition would also change
on similar timescales.
In future work, we shall study the long-term dispersal process of a PPD by incorporating the formation, growth, and destruction of dust grains.

\acknowledgments
The authors thank Riouhei Nakatani for
discussion on the disk chemo-thermal structure.  
Numerical computations were carried out on Cray XC50 at Center for Computational Astrophysics, National Astronomical Observatory of Japan.
%
\vspace{5mm}





\bibliography{bibliography}
\bibliographystyle{aasjournal}



\end{document}